\documentclass[jcp,reprint,floatfix,aip]{revtex4-1}

\usepackage{bm}
\usepackage{float}
\usepackage{graphicx}
\usepackage{color}
\usepackage{dcolumn} 
\definecolor{cfb}{rgb}{0.0,0.9,0.9}
\definecolor{orange}{rgb}{1.0,0.45,0.0}
\definecolor{lila}{rgb}{0.7,0.0,1.0}

 \usepackage{color}
 \usepackage[normalem]{ulem}
 \usepackage{cancel}
\begin{document}

\title{Four-Component Density Matrix Renormalization Group}

\author{Stefan Knecht}
\email{stefan.knecht@phys.chem.ethz.ch}
\affiliation{ETH Z\"urich, 
Laboratory of Physical Chemistry, 
Vladimir-Prelog-Weg 2, 8093 Z\"urich, Switzerland}

\author{\"Ors Legeza}
\email{legeza.ors@wigner.mta.hu}
\affiliation{
Strongly Correlated Systems ``Lend\"ulet" Research Group, 
Wigner Research Center for Physics, H-1525 Budapest, Hungary
}

\author{Markus Reiher}
\email{markus.reiher@phys.chem.ethz.ch}
\affiliation{ETH Z\"urich, 
Laboratory of Physical Chemistry, 
Vladimir-Prelog-Weg 2, 8093 Z\"urich, Switzerland}

\begin{abstract}
We present the first implementation of the relativistic quantum chemical two- and four-component density matrix 
renormalization group algorithm that includes a variational description of scalar-relativistic effects and spin--orbit coupling. 
Numerical results based on the four-component Dirac--Coulomb Hamiltonian are presented for the standard reference molecule for correlated relativistic benchmarks:
thallium hydride. 
\end{abstract}

\maketitle


Owing to remarkable advances in the past decades relativistic quantum chemical methods have become  
a routinely applicable and indispensable tool for the accurate description of the chemistry and spectroscopy 
of molecular compounds comprising heavy elements \cite{perspectives_rel_qc12,dyall_faegri_qc,reiher_qc}. 
Also the complete understanding of the photochemistry and photophysics of first- and second-row molecules requires to encompass 
relativistic effects --- the most important of which are spin--orbit (SO) interactions --- needed to calculate   
intersystem crossing rates \cite{WCMS:WCMS83}.  
Major challenges for relativistic quantum chemistry originate from (i)\ the reduction of non-relativistic (spin and spatial) symmetries caused by magnetic couplings that lead to in general complex wave functions and require the use of 
double-group symmetry 
as well as (ii)\ the large number of (unpaired) valence electrons to be correlated (in particular for heavy elements)  
and (iii) the occurrence of near-degeneracies of electronic states. 
Popular quantum chemical methods such as CASSCF/CASPT2/SO-CASPT2 \cite{caspt2-review-wires2012}\ 
assume an additivity of electron
correlation and spin--orbit effects or a weak polarization of orbitals due to spin--orbit interaction, or both.
Hence, for heavy-element compounds accuracy is inevitably limited as relativistic effects and static or dynamic electron correlation 
are often not only large but also counteracting \cite{pyykkoe88,perspectives_rel_qc12}. 

To address the latter issue adequately, a number of genuine relativistic multiconfigurational and multireference 
approaches have been proposed \cite{Fleig20122-review-rel-correlation-methods,neese_socasscf12,Kim2013}. 
In this Communication we merge the strengths of the density matrix renormalization group (DMRG) algorithm \cite{white,White1999}, which 
has been successfully introduced to the field of non-relativistic quantum chemistry \cite{ors_springer,reiher10_dmrg,chanreview}, 
with a variational description of all relativistic effects in the orbital basis. This new four-component (4c) DMRG \emph{ansatz}\ goes beyond 
preceeding scalar-relativistic DMRG approaches \cite{C3CP53975J,Moritz2005a} and allows us to efficiently describe first and foremost non-dynamic correlation (or strong correlations) in heavy-element complexes by means of extensive active orbital spaces 
which would surmount capabilities of any to-date available relativistic multiconfigurational approach. 

The point of departure for our relativistic DMRG implementation is the time-independent
4c-Dirac--Coulomb(--Breit) Hamiltonian \cite{saue_primer11} --- any suitable (exact) two-component (2c) Hamiltonian approximation is directly amendable, too. The basic four-component electronic eigenvalue equation for a many-particle system is conveniently cast (with positive-energy projectors omitted) in a form which is known from non-relativistic quantum chemistry \cite{dyall_faegri_qc,reiher_qc},
\begin{equation}
\label{rel-ham}
\hspace{-0.30cm} \hat{H} \Psi = \left(\sum\limits_{i}\hat{h}_D(i)+\frac{1}{2} \sum\limits_{i \neq j}\hat{g}(i,j)+V_{NN}\right) \Psi = E_{\rm el} \Psi ,
\end{equation}
where $\hat{h}_D(i)$\ is the one-electron Dirac Hamiltonian for electron $i$, $\hat{g}(i,j)$\ is a two-electron operator describing the interaction between electrons $i$\ and $j$, $V_{NN}$\ is the classical nuclear repulsion energy operator, $E_{el}$\ is the energy eigenvalue and $\Psi$\ is a four-component wave function. 
In the absence of any external magnetic field it can be shown that Eq. (\ref{rel-ham})\ is symmetric under time-reversal \cite{reiher_qc}\ from which follows that a fermion four-component spinor functions $\phi_i$ occurs in Kramers pairs $\{\phi_i,\bar{\phi}_i\}$. A spinor $\bar{\phi}_i$\ can thus be obtained from the action of the time-reversal operator $\hat{\mathcal{K}} = -i\Sigma_y \hat{\mathcal{K}}_0$\ on ${\phi}_i$, that is ${\hat\mathcal{K}}\phi_i = \bar{\phi}_i$. Hence, our 4c- (or 2c-)spinor basis is comprised of Kramers pairs\ which we will imply in the following derivations.
 
In the no-pair approximation, we can formulate the resulting Hamiltonian in second-quantized and normal ordered form,
\begin{equation}
\hat{\widetilde{H}} = \sum\limits_{PQ}^{}\ F_{P}^{Q} \{a_{{P}}^{\dagger}a_Q\} + \frac{1}{4} \sum\limits_{PQRS}^{}\ V_{PR}^{QS} \{a_{{P}}^{\dagger}a_{R}^{\dagger}a_{S}a_{Q}\}\ ,
\label{rel-h-2nd-quant}
\end{equation} 
where the summation indices $PQRS$\ strictly refer to positive-energy orbitals, and $F_{P}^{Q}$\ and $V_{PR}^{QS} = (G_{PR}^{QS} - G_{PR}^{SQ} ) $\ are Fock-matrix elements and antisymmetrized two-electron integrals\ $G_{PR}^{QS}$, respectively. The Hamiltonian $\hat{\widetilde{H}}$\ in Eq. (\ref{rel-h-2nd-quant})\ constitutes the starting point for our 4c-DMRG implementation. 

We benefit from a quaternion symmetry scheme \cite{saue1999_quaternion-symmetry-dirac}\ that has been implemented for the binary double groups D$_{2h}^{\ast}$\ and subgroups thereof in the \texttt{Dirac}\ 
program package \cite{DIRAC12} to which our DMRG program is interfaced. 
In this scheme, point group symmetry and quaternion operator algebra are combined advantageously such that the eigenvalue equation, Eq.~(\ref{rel-ham}), 
can be solved either using real (double groups D$_{2h}^{\ast}$, D$_{2}^{\ast}$\ and C$_{2v}^{\ast}$; resulting number of non-zero real matrices of a quaternion operator matrix representation: NZ = 1), complex (C$_{2h}^{\ast}$, C$_{2}^{\ast}$\ and C$_{s}^{\ast}$; NZ = 2) 
or quaternion algebra (C$_{i}^{\ast}$\ and C$_{1}^{\ast}$; NZ = 4). Working in a Kramers-paired spinor basis, one can then show that all operator matrix elements $t_{p\bar{q}}$\ of a time-symmetric one-electron operator $\hat{t}$\ are zero by symmetry. Furthermore the complete set of two-electron integrals $G_{PR}^{QS}$\ 
of the two-electron (Coulomb) operator $\hat{g}$\ in molecular orbital (MO) basis can be cast 
into a $4 \times 3$\ ((NZ,3))\ matrix representation (see also Appendix B.3 page 161\emph{ff} of Ref. \cite{thyssendiss}),
\begin{eqnarray}
\mathbf{{G}} = 
\left(\begin{array}{ccc}
\mathcal{R}((PQ|RS)) & \mathcal{R}((P\bar{Q}|R\bar{S})) & \mathcal{R}((\bar{P}Q|\bar{R}S)) \\
\mathcal{I}((PQ|RS)) & \mathcal{I}((P\bar{Q}|R\bar{S})) & \mathcal{I}((\bar{P}Q|\bar{R}S)) \\
\mathcal{R}((PQ|R\bar{S})) & \mathcal{R}((P\bar{Q}|R{S})) & \mathcal{R}((\bar{P}Q|{R}S)) \\
\mathcal{I}((PQ|R\bar{S})) & \mathcal{I}((P\bar{Q}|R{S})) & \mathcal{I}((\bar{P}Q|{R}S)) 
\end{array} \right), 
\label{dirac:nz-3-format}
\end{eqnarray}
where $\mathcal{R}$\ and $\mathcal{I}$\ denote the real and complex parts of a two-electron integral in MO representation, respectively, and $P,Q,R,S$\ label spinor indices. 
The number of nonzero rows for a given binary double group thus corresponds to the NZ rank as given above. 
Important symmetry reductions for both the one- and two-electron integrals are therefore being taken into account in a relativistic Kramers-unrestricted DMRG implementation. This scheme not only provides considerable computational savings but also ensures that the DMRG wave function has the correct time-reversal symmetry in case of a closed-shell molecule. 

In a Kramers-restricted spinor basis all one-electron matrix elements $F_{P}^{Q}$\ (see Eq. \ref{rel-h-2nd-quant}) among 
barred and unbarred components will be identical while matrix elements between barred 
and unbarred are non-zero only in the NZ=4 case. In contrast, a two-electron integral $G_{PR}^{QS}$\ may generally 
be comprised of barred and unbarred spinors. As illustrated by Eq.~(\ref{dirac:nz-3-format})\ for NZ=1 and NZ=2, respectively, only an even number ($n_{\rm barred}=0,2,4$) of barred spinors yields a non-vanishing two-electron integral whereas for NZ=4 all combinations are contributing.  
Even though integrals can be made real-valued (NZ=1), permutational symmetry is reduced by a factor two compared to the 8-fold permutational symmetry in the non-relativistic case since orbitals are complex in a relativistic framework. 

In DMRG, electron--electron correlation
is taken into account by an iterative
procedure that minimizes the Rayleigh quotient corresponding
to the electronic Hamiltonian $\hat{\widetilde{H}}$\ and eventually converges a full-CI-type wave function within the selected active orbital space.
The full configuration Hilbert space of a finite system comprising $N$ MOs, $\Lambda^{(N)}$,
is built from tensor product spaces of local orbital (tensor) spaces $\Lambda_i$,\cite{new_review} which can be written as
$\Lambda^{(N)}=\otimes_{i=1}^N \Lambda_i$.
Since standard non-relativistic DMRG implementations usually employ a spatial-orbital basis, 
the dimension of the local Hilbert space of a single molecular orbital, $q={\rm dim}\ \Lambda_i$, becomes 4
while the full dimensionality is ${\rm dim}\ \Lambda^{(N)}=4^N$. In this representation an MO can be either empty, 
singly occupied with spin up or down, or doubly occupied with paired spins. 
Our implementation exploits a two-dimensional local Hilbert-space representation, $q=2$, where each spinor can either be empty or singly occupied. The tensor space dimension is then $2^{N}$ with $N$ being the number of spinors.

In the two-site DMRG variant \cite{white}, that is the basis for our relativistic DMRG implementation, $\Lambda^{(N)}$ is approximated by a tensor product space of four tensor spaces, i.e., 
$\Lambda^{(N)}_{\rm DMRG}=\Lambda^{(l)}\otimes\Lambda_{l+1}\otimes\Lambda_{l+2}\otimes\Lambda^{(r)}$. The dimensions of the corresponding local \emph{left} ($l$) \ and \emph{right}\ ($r$)\ spaces are denoted as $M_l={\rm \dim}\ \Lambda^{(l)}$ and $M_r={\rm \dim}\ \Lambda^{(r)}$, respectively. 
With $q={\rm \dim}\ \Lambda_{l+1}={\rm \dim}\ \Lambda_{l+2}$\ the resulting dimensionality of the DMRG wave function is 
${\rm \dim}\ \Lambda^{(N)}_{\rm DMRG}=q^2M_lM_r\ll q^N$.
The number of block states, $M_l$ and $M_r$, required to achieve sufficient convergence can be regarded as a function of the level of entanglement among the molecular orbitals. Hence the maximum number of block states $M_{\rm max} = \max{(M_l,M_r)}$\ determines 
the accuracy of a DMRG calculation \cite{legeza_dbss2}. 

The success and numerical efficiency of the DMRG algorithm 
rely on a subsequent application of the singular value decomposition (SVD) theorem \cite{Schollwock2010,new_review} while the performance depends on the level of entanglement encoded in the wave function \cite{legeza_dbss3}.
During an SVD step, the finite system is divided into two parts by 
expressing $\Lambda^{(N)}_{\rm DMRG}=\Lambda^{(L)}\otimes\Lambda^{(R)}$, namely the system and environment blocks, 
where $\Lambda^{(L)}=\Lambda^{(l)}\otimes\Lambda_{l+1}$, and 
$\Lambda^{(R)}=\Lambda_{l+2}\otimes\Lambda^{(r)}$. 
In each DMRG step, the basis states of the system block are then transformed to a new {\em truncated basis} set 
by a unitary transformation based on the preceeding SVD\cite{schollwoeck}. This transformation  
depends therefore on how accurately the environment is represented \cite{Moritz2006}\ as well as on the level of truncation\cite{legeza_dbss2}. 
As a consequence the accuracy of the DMRG method is governed by the truncation error, $\delta \varepsilon_{\rm TR}$, as well as by the 
environmental error, $\delta \varepsilon_{\rm sweep}$ \cite{legeza1996}. The latter is minimized in each DMRG macro-iteration by a successive application of the SVD going through the system back and forth (``sweeping"). 

In order to minimize $\delta \varepsilon_{\rm sweep}$, which is usually largest during the initial sweep of the DMRG approach 
because of a poor representation of the environment, we take advantage of the Configuration Interaction based
Extended Active Space procedure (CI-DEAS) \cite{legeza_dbss,legeza2004-leiden} to efficiently construct 
the environmental basis states by means of an orbital entropy profile \cite{dmrg-quantum-information-analysis-2011}. 
The latter is dependent on the orbital ordering along a (fictitious) one-dimensional chain \cite{legeza_dbss3,dmrggerrit3}\ and determines the maximum number of block states $M_{\rm max}=\max(M)$ 
that is needed to satisfy an \emph{a priori} defined accuracy threshold given by a value $\chi$. 

The truncation error $\delta \varepsilon_{\rm TR}$\ is a function of the total number of block states $M$.
Assuming $M_l=M_r=M$\ we can exploit a second-order polynomial fit as a function of $1/M$ by taking the limit\ of zero energy change between two sweeps $E(M, \delta \varepsilon_{\rm sweep}=0)$\ 
for a given $M$\ to provide a good estimate for the truncation-free solution \cite{legeza1996,barcza2012}. 

We demonstrate the capabilities of our 4c-DMRG implementation at the example of 
the thallium hydride molecule since this system has become a standard benchmark molecule for a plethora of 
relativistic methods  \cite{seijo1995_tlh,lenthe1996_tlh,han1999_tlh,titov2000_tlh,visscher2001_tlh_ccsdt,mayer2001_tlh,choi2001_tlh,ilias2001_tlh,choi2003_tlh,zeng2010-tlh-so}\ (see also references in Ref.~\citenum{zeng2010-tlh-so}). 
Orbitals and MO integrals were computed  
with a development version of the \texttt{Dirac12}\ program package \cite{DIRAC12}\ using the Dirac--Coulomb (DC) Hamiltonian and 
triple-$\zeta$\ basis sets for Tl (cv3z) \cite{dyall_6p_all,dyall_6p_core_corr2012} and H (cc-pVTZ) \cite{dunning89}, 
which include core-correlating functions for Tl. All DMRG calculations were performed with 
the relativistic development branch of the \textsc{QC-DMRG-Budapest}\ program \cite{Legeza}. 
C$_{2v}^{\ast}$\ double group symmetry (NZ=1) was assumed throughout all calculations for TlH.
MP2 natural spinors (NSs)\ \cite{mp2-no}, correlating the Tl $5s5p4f5d6s6p$\ and H $1s$\ electrons while keeping the remaining core electrons of Tl frozen, served as the orbital basis for all electron-correlation calculations.  
Since \texttt{Dirac12}\ requires to use uncontracted basis sets in a four-component framework, a virtual orbital threshold was set at 135 hartree, such that the initial virtual correlation space in the MP2 calculation 
comprised all recommended core-valence and valence-correlation functions. 
The final active space was then chosen to include all occupied spinors that have MP2-NS occupancies less than $1.98$\ 
as well as all virtuals up to a cutoff of $\approx 0.001$\ in the MP2-NS occupation numbers. 
Given this criterion, an active space of 14 electrons --- the occupied Tl $5d6s6p$ plus H $1s$\ shells --- in 47 Kramers pairs (94 spinors) was used in the CI \cite{fleig03,fleig06,knecht10a}\, MP2, CC  \cite{mrcceng,Kallay_2001,Kallay_2001}\ and DMRG calculations. 
The latter are further characterized by the choice of $M_{\rm min}, M_{\rm max}, M_{\rm min}^{\rm DEAS}$\ and $\chi$, denoted in the following as DMRG(14,94)[$M_{\rm min},M_{\rm max},M_{\rm min}^{\rm DEAS},\chi$]. 

\begin{figure}[H]
\centerline{
\includegraphics[scale=0.375]{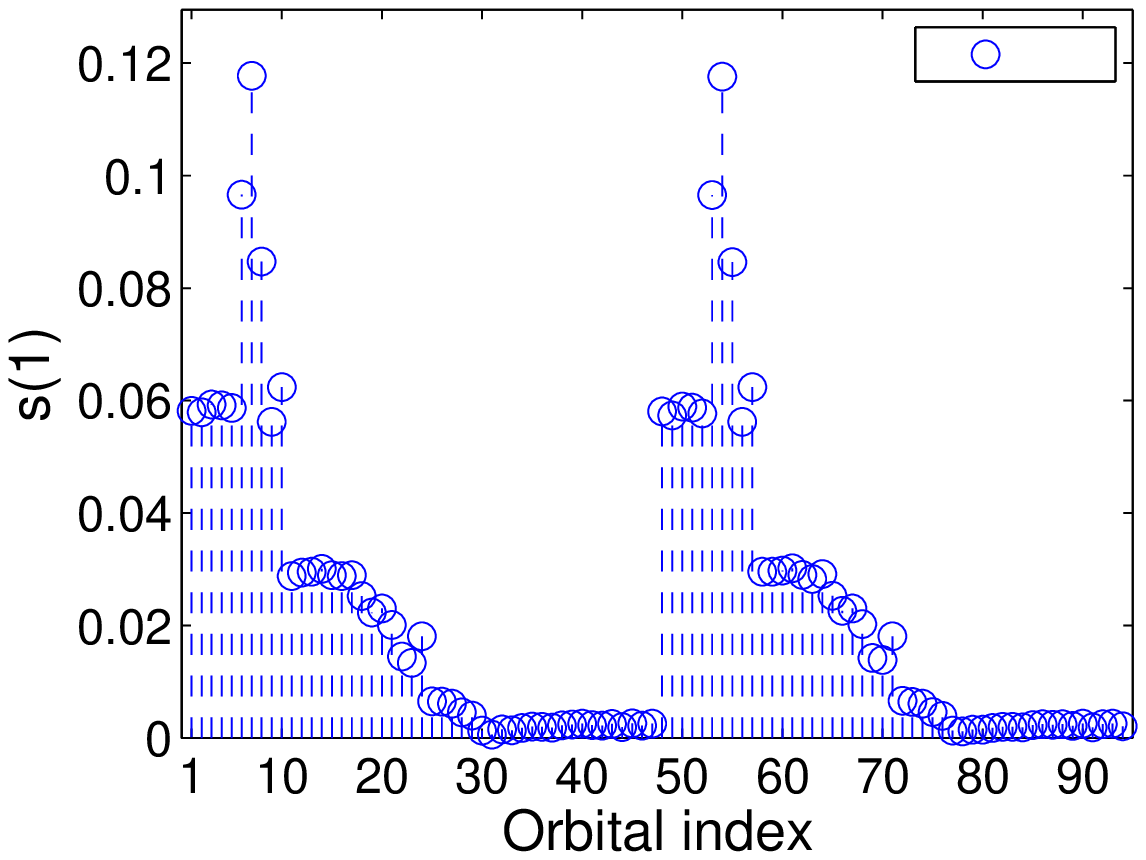}
\includegraphics[scale=0.375]{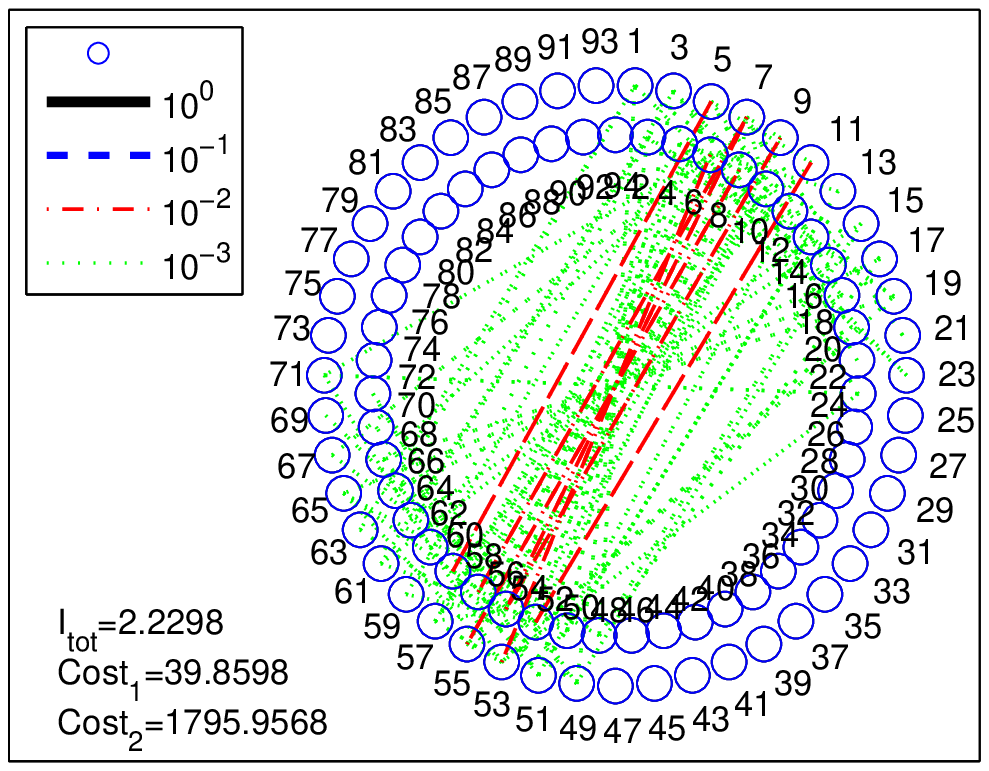}
}
\caption{\label{fig:entropy}Left: One-orbital entropy profile, $s_i$, calculated at the experimental internuclear distance $r_e^{\rm exp}$=1.872 \AA. The larger the entropy value for a given spinor the larger its contribution to the total correlation energy. Right: Schematic plot of a piecewise orbital entanglement based on the two-orbital mutual information, $I_{ij}$. Entanglement strengths are indicated by different colors.}
\end{figure}

Figure~\ref{fig:entropy} depicts the one- $s_i$ and two-orbital $I_{ij}$\ entropy profiles \cite{legeza_dbss3,dmrg-entanglement12,boguslawski2013a} at the experimental internuclear distance $r_e^{\rm exp}$=1.872 \AA\   computed from an initial DMRG[256,256,256,10$^{-5}$] calculation. We first note that the 
one-orbital entropy profile, (left-hand side of Figure~\ref{fig:entropy}) is nearly perfectly symmetric with respect to the unbarred (\#1--\#47) and barred (\#48--\#94) spinors where any slight deviation is an artefact of the preset low $M_{\rm min},M_{\rm max}$\ values. 
The total quantum information $I_{\rm tot}$\ encoded in the wave function, defined as the sum of one-orbital entropies,
$I_{\rm tot}=\sum_{i} s_i$, 
can be taken as a measure of the 
importance of dynamic (weak) electron correlation. The lower $I_{\rm tot}$\ (compared to 
$I_{\rm tot}^{\rm max}=\sum_i s_i^{\rm max}= N\ln(2)=65.15$), the more important will be an appropriate account of 
dynamic electron correlation in order to grasp all important correlation effects.
In the present case of TlH we have $I_{\rm tot}\simeq 2.23\ll I_{\rm tot}^{\rm max}$\ which
points to the fact that TlH is a predominantly single-reference close to its equilibrium structure.  

The two-orbital mutual information, $I_{ij}$, confirms this qualitative picture. $I_{ij}$ values are visualized in the right panel of Figure~\ref{fig:entropy}, 
where the degree of entanglement between spinors is marked by a color-coded connecting line.  
While few spinors are weakly entangled (red)  
the majority is entangled with even smaller strengths (green). 
Since several spinors are mutually entangled with the same order of magnitude, we expect that large $M_{\rm min}, M_{\rm max}$\ 
values combined with a low quantum information loss 
threshold $\chi$\ are required to reach a fully converged DMRG wave function. 

\begin{figure}[!]
\centerline{
\includegraphics[scale=0.58]{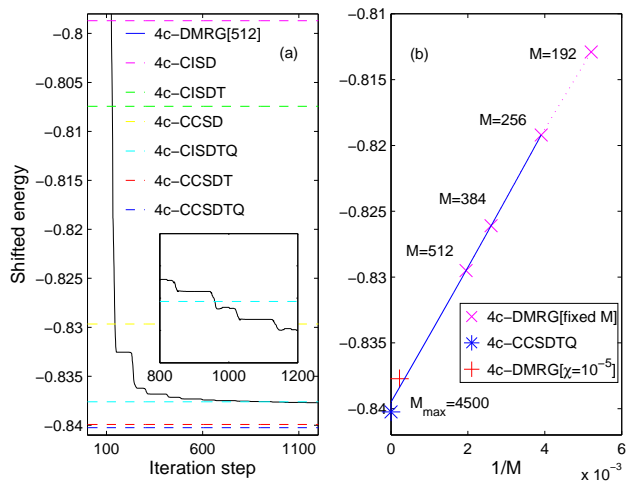}
}
\caption{Left: Convergence of the ground state energy (shifted by 20275 H) as a function of iteration steps of the 4c-DMRG(14,94)[4500,1024,2048,10$^{-5}$]\ approach at $r_e^{\rm exp}$=1.872 \AA. 
Reference energies calculated by various CI and CC wave function models are also given as horizontal lines. 
The inset shows that the 4c-DMRG energy drops below the 4c-CI-SDTQ energy.
Right: Extrapolation of DMRG energies E($M$)-20275 H\ for fixed $M$\ values towards the limit $E(M\rightarrow\infty)$-20275 H.
}
\label{fig:energy}
\end{figure}

To corroborate this hypothesis we compiled in Table \ref{tlh-results}\ total energy differences for various standard wave-function-expansion
methods as well as for our 4c-DMRG(14,94)[4500,1024,2048,10$^{-5}$]\ model with respect to a chosen 4c-CCSDTQ reference at $r_e^{\rm exp}$=1.872 \AA. The 4c-DMRG wave function was built from an optimized ordering of orbitals based on the entropy profiles given in Figure~\ref{fig:entropy}\ and by applying high accuracy settings in the initial CI-DEAS sweep (with ${\rm CI}_{\rm level}=4$\ and $\chi_{\rm CI}=10^{-8}$). These initial conditions ensured both a rapid elimination of the environmental error 
and a fast total convergence towards the global minimum as illustrated by the left-hand side of Figure~\ref{fig:energy}. 
The 4c-CISDTQ energy is in fact reached after no more than six sweeps of the 4c-DMRG wave function optimization procedure.  
\begin{table}[!]
\begin{ruledtabular}
  \caption{Total electronic energy differences $\Delta$E$_{\rm el}$ (in mH) for different correlation approaches with respect to the 4c-CCSDTQ(14,94) reference energy of -20275840.24233 mH for TlH computed at the experimental equilibrium internuclear distance 1.872 \AA.}
 \begin{tabular}{lr} \noalign{\smallskip}
  method & $\Delta$E$_{\rm el}$\\ \hline
    4c-CISD(14,94)           &   41.55 \\
    4-CISDT(14,94)         &   32.80 \\
    4c-CISDTQ(14,94)      &   2.63 \\
    4c-MP2(14,94)            &  -13.49\\
    4c-CCSD(14,94)         &  10.58\\
    4c-CCSD(T)(14,94)     &  -0.32\\
    4c-CCSDT(14,94)       &   0.33\\
    4c-CCSDT(Q)(14,94)  &  -0.07\\
    4c-DMRG(14,94)[4500,1024,2048,10$^{-5}$]        &    2.57\\ 
    4c-DMRG(14,94)[$M\rightarrow\infty$ extrapolated]       &   0.7 
\end{tabular} 
\label{tlh-results}
\end{ruledtabular}
\end{table}
Inspection of Table \ref{tlh-results}\ furthermore reveals that 
the 4c-DMRG energy is, although being below our best variational 4c-CISDTQ energy, 
still 2.57 mH higher than the reference 4c-CCSDTQ as well as 2.89 mH higher than the single-reference 4c-CCSD(T) energies. 
From the convergence pattern of our benchmark 4c-DMRG[4500,1024,2048,10$^{-5}$] calculation displayed on the left-hand side of Figure~\ref{fig:energy}\ the following picture emerges: 
after having reached the maximum number of block states $M_{\rm max}$ ($\approx$\ 3 sweeps) ---  for the present problem we have a computational limit of $M_{\rm max}=4500$ --- the convergence rate slows down significantly and after the sixth sweep the energy is no longer a decreasing function of the iteration steps because the environmental error now starts to fluctuate to a certain extent depending on the actual superblock configuration. As a result the 4c-CCSDTQ reference energy could not be reached (see Table \ref{tlh-results}), which is, however, not a fundamental problem of the approach. 
It must be emphasized again that DMRG is best suited for static-correlation problems while TlH is dominated by dynamic correlation, for which CC approaches are much more suitable.  
However, extrapolating the DMRG energy for a given $M$ to the limit\ $E(M, \delta \varepsilon_{\rm sweep}=0)$\ by using an exponential function in $1/\rm{sweep}$ (\emph{vide supra})\ provides an effective means to eliminate the truncation error. The {right-hand side} of Figure~\ref{fig:energy}\ shows the extrapolated energies
along with the resulting best-estimate DMRG energy $E(M\rightarrow\infty)=-20275.8395$\ H. At $r_e^{\rm exp}$=1.872 \AA\  
$E(M\rightarrow\infty)$\ is now as close as +0.7 mH to the CCSDTQ reference energy. 
Taking further advantage of the extrapolation scheme we exploit a characteristic feature of the DMRG parametrization, namely, that it includes, in contrast to a CI expansion, all excitations required to describe the wave function to a given accuracy. This property implies that the general structure
of the DMRG wave function is preserved and can be determined even with smaller
$M$\ values \cite{legeza_dbss2}. 
We found that the resulting 4c-DMRG[512, $\delta \varepsilon_{\rm sweep}=0$] potential energy curve does not 
only effectively reproduce the shape of the 4c-CCSDTQ potential energy curve but also yields accurate 
spectroscopic constants --- compiled in Table~\ref{tlh-spec-results} --- as extracted from a fourth-order polynomial fit. 
The 4c-CCSDTQ data is in excellent agreement with experiment for the equilibrium internuclear distance $r_e$, harmonic frequency $\omega_e$, and for the anharmonicity constant $\omega_e x_e$\ while it turns out to be superior to a selection of other theoretical approaches listed in Table~\ref{tlh-spec-results}. The current DMRG results 
in turn show an excellent agreement with experiment for $r_e$\ while predicting slightly too high values for $\omega_e$\ (+20 cm$^{-1}$) 
and $\omega_e x_e$ ($+4$ cm$^{-1}$), respectively. 

We conclude with a note on the computational demands of our 4c-DMRG implementation 
in comparison to CCSDTQ. The benchmark DMRG[4500,1024,2048,10$^{-5}$] calculation 
required $\approx$ 50 GB of core memory to represent all operators of the left and right blocks 
while the relativistic MRCC code \cite{kallay2011_4cCC}\ had similar memory requirements 
for the optimization of the various $t$\ amplitudes.

Our new 4c- and 2c-DMRG approach (no 2c-results are shown here) bears the potential to become a new reference approach comparable to higher-order CC methods, in particular for molecular systems that exhibit rather strong multi-configurational character in their ground- and excited states. To improve on the description of dynamic correlation, a further combination with standard approaches like, for example, multi-reference perturbation theory is possible and will yield a powerful tool for the theoretical chemistry and photophysics of heavy-element molecules like lanthanide and actinide complexes.

\begin{table}[!]
  \caption{Spectroscopic constants of $^{205}$TlH obtained from 4c-DMRG[$512,\delta \varepsilon_{\rm sweep}=0)$], CI and CC calculations in comparison with other 
  theoretical and experimental work.}
\begin{ruledtabular}
 \begin{tabular}{lccc}
  method & r$_e$ [\AA] & $\omega_e$ [cm$^{-1}$] &  $\omega_e x_e$ [cm$^{-1}$]  \\ \hline
  4c-DMRG(14,94)[512]      &1.873 & 1411 & 26.64 \\
  4c-CISD(14,94)          & 1.856 & 1462  & 23.11  \\
  4c-CISDTQ(14,94)      & 1.871 & 1405  & 20.11  \\
  4c-MP2(14,94)            & 1.828 & 1546 & 47.27  \\
  4c-CCSD(14,94)          & 1.871 & 1405 & 19.36  \\
  4c-CCSD(T)(14,94)      & 1.873 & 1400 & 23.52  \\
  4c-CCSDT(14,94)        & 1.873 & 1398 & 22.28  \\
  4c-CCSDT(Q)(14,94)   & 1.873 & 1397 & 21.01  \\
  4c-CCSDTQ(14,94)     & 1.873 & 1397 & 22.24 \\
  CCSD(T)$^a$        & 1.876 & 1385 & n/a \\
  CCSD(T)$^b$        & 1.877 & 1376 & n/a \\
  MRD-CI$^c$         & 1.870 & 1420 & n/a  \\
  SO-MCQDPT$^d$ & 1.876 & 1391 & 29.42  \\
 experiment$^e$     & 1.872 & 1390.7 & 22.7  \\ 
\end{tabular} 
\end{ruledtabular}
\begin{flushleft}
 $^a$\ 4c-DC CCSD(T) [14 electrons], see Ref. \citenum{visscher2001_tlh_ccsdt}.\\
 $^b$\ 4c-DC-Gaunt CCSD(T) [36 electrons], see Ref. \citenum{visscher2001_tlh_ccsdt}.\\
 $^c$\ GRECP spin--orbit MRD-CI, see Ref. \citenum{titov2000_tlh}.\\
 $^d$\ model-core potential spin--orbit MCQDPT, see Ref. \citenum{zeng2010-tlh-so}.\\
 $^e$\ experimental data taken from Refs. \citenum{tlh_experiment_I_1938,huber79,Urban1989_tlh_experiment_II,titov2000_tlh}.
\end{flushleft}
\label{tlh-spec-results}
\end{table}

This work has been supported financially in part by the Schweizerischer Nationalfonds (project no. 200020\_144458/1)
and by the Hungarian Research Fund (OTKA) through Grants No. K100908 and No. NN110360.


\begin{thebibliography}{58}%
\makeatletter
\providecommand \@ifxundefined [1]{%
 \@ifx{#1\undefined}
}%
\providecommand \@ifnum [1]{%
 \ifnum #1\expandafter \@firstoftwo
 \else \expandafter \@secondoftwo
 \fi
}%
\providecommand \@ifx [1]{%
 \ifx #1\expandafter \@firstoftwo
 \else \expandafter \@secondoftwo
 \fi
}%
\providecommand \natexlab [1]{#1}%
\providecommand \enquote  [1]{``#1''}%
\providecommand \bibnamefont  [1]{#1}%
\providecommand \bibfnamefont [1]{#1}%
\providecommand \citenamefont [1]{#1}%
\providecommand \href@noop [0]{\@secondoftwo}%
\providecommand \href [0]{\begingroup \@sanitize@url \@href}%
\providecommand \@href[1]{\@@startlink{#1}\@@href}%
\providecommand \@@href[1]{\endgroup#1\@@endlink}%
\providecommand \@sanitize@url [0]{\catcode `\\12\catcode `\$12\catcode
  `\&12\catcode `\#12\catcode `\^12\catcode `\_12\catcode `\%12\relax}%
\providecommand \@@startlink[1]{}%
\providecommand \@@endlink[0]{}%
\providecommand \url  [0]{\begingroup\@sanitize@url \@url }%
\providecommand \@url [1]{\endgroup\@href {#1}{\urlprefix }}%
\providecommand \urlprefix  [0]{URL }%
\providecommand \Eprint [0]{\href }%
\providecommand \doibase [0]{http://dx.doi.org/}%
\providecommand \selectlanguage [0]{\@gobble}%
\providecommand \bibinfo  [0]{\@secondoftwo}%
\providecommand \bibfield  [0]{\@secondoftwo}%
\providecommand \translation [1]{[#1]}%
\providecommand \BibitemOpen [0]{}%
\providecommand \bibitemStop [0]{}%
\providecommand \bibitemNoStop [0]{.\EOS\space}%
\providecommand \EOS [0]{\spacefactor3000\relax}%
\providecommand \BibitemShut  [1]{\csname bibitem#1\endcsname}%
\let\auto@bib@innerbib\@empty
\bibitem [{\citenamefont {Autschbach}(2012)}]{perspectives_rel_qc12}%
  \BibitemOpen
  \bibfield  {author} {\bibinfo {author} {\bibfnamefont {J.}~\bibnamefont
  {Autschbach}},\ }\href@noop {} {\bibfield  {journal} {\bibinfo  {journal}
  {J.\,Chem.\,Phys.}\ }\textbf {\bibinfo {volume} {136}},\ \bibinfo {pages}
  {150902} (\bibinfo {year} {2012})}\BibitemShut {NoStop}%
\bibitem [{\citenamefont {Dyall}\ and\ \citenamefont
  {F{\ae}gri}(2007)}]{dyall_faegri_qc}%
  \BibitemOpen
  \bibfield  {author} {\bibinfo {author} {\bibfnamefont {K.~G.}\ \bibnamefont
  {Dyall}}\ and\ \bibinfo {author} {\bibfnamefont {K.}~\bibnamefont
  {F{\ae}gri}},\ }\href@noop {} {\emph {\bibinfo {title} {Introduction to
  Relativistic Quantum Chemistry}}}\ (\bibinfo  {publisher} {Oxford University
  Press},\ \bibinfo {address} {Oxford},\ \bibinfo {year} {2007})\BibitemShut
  {NoStop}%
\bibitem [{\citenamefont {Reiher}\ and\ \citenamefont
  {Wolf}(2009)}]{reiher_qc}%
  \BibitemOpen
  \bibfield  {author} {\bibinfo {author} {\bibfnamefont {M.}~\bibnamefont
  {Reiher}}\ and\ \bibinfo {author} {\bibfnamefont {A.}~\bibnamefont {Wolf}},\
  }\href@noop {} {\emph {\bibinfo {title} {Relativistic Quantum Chemistry: The
  Fundamental Theory of Molecular Science}}}\ (\bibinfo  {publisher}
  {Wiley-VCH},\ \bibinfo {address} {Weinheim},\ \bibinfo {year}
  {2009})\BibitemShut {NoStop}%
\bibitem [{\citenamefont {Marian}(2012)}]{WCMS:WCMS83}%
  \BibitemOpen
  \bibfield  {author} {\bibinfo {author} {\bibfnamefont {C.~M.}\ \bibnamefont
  {Marian}},\ }\href {\doibase 10.1002/wcms.83} {\bibfield  {journal} {\bibinfo
   {journal} {{WIREs} Comp. Mol. Sci.}\ }\textbf {\bibinfo {volume} {2}},\
  \bibinfo {pages} {187} (\bibinfo {year} {2012})}\BibitemShut {NoStop}%
\bibitem [{\citenamefont {Roca-Sanjuan}, \citenamefont {Aquilante},\ and\
  \citenamefont {Lindh}(2012)}]{caspt2-review-wires2012}%
  \BibitemOpen
  \bibfield  {author} {\bibinfo {author} {\bibfnamefont {D.}~\bibnamefont
  {Roca-Sanjuan}}, \bibinfo {author} {\bibfnamefont {F.}~\bibnamefont
  {Aquilante}}, \ and\ \bibinfo {author} {\bibfnamefont {R.}~\bibnamefont
  {Lindh}},\ }\href {\doibase 10.1002/wcms.97} {\bibfield  {journal} {\bibinfo
  {journal} {{WIREs} Comp. Mol. Sci.}\ }\textbf {\bibinfo {volume} {2}},\
  \bibinfo {pages} {585} (\bibinfo {year} {2012})}\BibitemShut {NoStop}%
\bibitem [{\citenamefont {Pyykk{\"o}}(1988)}]{pyykkoe88}%
  \BibitemOpen
  \bibfield  {author} {\bibinfo {author} {\bibfnamefont {P.}~\bibnamefont
  {Pyykk{\"o}}},\ }\href@noop {} {\bibfield  {journal} {\bibinfo  {journal}
  {Chem. Rev.}\ }\textbf {\bibinfo {volume} {88}},\ \bibinfo {pages} {563}
  (\bibinfo {year} {1988})}\BibitemShut {NoStop}%
\bibitem [{\citenamefont
  {Fleig}(2012)}]{Fleig20122-review-rel-correlation-methods}%
  \BibitemOpen
  \bibfield  {author} {\bibinfo {author} {\bibfnamefont {T.}~\bibnamefont
  {Fleig}},\ }\href {\doibase 10.1016/j.chemphys.2011.06.032} {\bibfield
  {journal} {\bibinfo  {journal} {Chem. Phys.}\ }\textbf {\bibinfo {volume}
  {395}},\ \bibinfo {pages} {2 } (\bibinfo {year} {2012})}\BibitemShut
  {NoStop}%
\bibitem [{\citenamefont {Ganyushin}\ and\ \citenamefont
  {Neese}(2013)}]{neese_socasscf12}%
  \BibitemOpen
  \bibfield  {author} {\bibinfo {author} {\bibfnamefont {D.}~\bibnamefont
  {Ganyushin}}\ and\ \bibinfo {author} {\bibfnamefont {F.}~\bibnamefont
  {Neese}},\ }\href@noop {} {\bibfield  {journal} {\bibinfo  {journal} {J.
  Chem. Phys.}\ }\textbf {\bibinfo {volume} {138}},\ \bibinfo {pages} {104113}
  (\bibinfo {year} {2013})}\BibitemShut {NoStop}%
\bibitem [{\citenamefont {Kim}\ and\ \citenamefont {Lee}(2013)}]{Kim2013}%
  \BibitemOpen
  \bibfield  {author} {\bibinfo {author} {\bibfnamefont {I.}~\bibnamefont
  {Kim}}\ and\ \bibinfo {author} {\bibfnamefont {Y.~S.}\ \bibnamefont {Lee}},\
  }\href {\doibase 10.1063/1.4822426} {\bibfield  {journal} {\bibinfo
  {journal} {J. Chem. Phys}\ }\textbf {\bibinfo {volume} {139}},\ \bibinfo
  {pages} {134115} (\bibinfo {year} {2013})}\BibitemShut {NoStop}%
\bibitem [{\citenamefont {White}(1992)}]{white}%
  \BibitemOpen
  \bibfield  {author} {\bibinfo {author} {\bibfnamefont {S.~R.}\ \bibnamefont
  {White}},\ }\href@noop {} {\bibfield  {journal} {\bibinfo  {journal} {Phys.
  Rev. Lett.}\ }\textbf {\bibinfo {volume} {69}},\ \bibinfo {pages} {2863}
  (\bibinfo {year} {1992})}\BibitemShut {NoStop}%
\bibitem [{\citenamefont {White}\ and\ \citenamefont
  {Martin}(1999)}]{White1999}%
  \BibitemOpen
  \bibfield  {author} {\bibinfo {author} {\bibfnamefont {S.~R.}\ \bibnamefont
  {White}}\ and\ \bibinfo {author} {\bibfnamefont {R.~L.}\ \bibnamefont
  {Martin}},\ }\href@noop {} {\bibfield  {journal} {\bibinfo  {journal} {J.
  Chem. Phys.}\ }\textbf {\bibinfo {volume} {110}},\ \bibinfo {pages} {4127}
  (\bibinfo {year} {1999})}\BibitemShut {NoStop}%
\bibitem [{\citenamefont {Legeza}\ \emph {et~al.}(2008)\citenamefont {Legeza},
  \citenamefont {Noack}, \citenamefont {S\'olyom},\ and\ \citenamefont
  {Tincani}}]{ors_springer}%
  \BibitemOpen
  \bibfield  {author} {\bibinfo {author} {\bibfnamefont {{\"O}.}~\bibnamefont
  {Legeza}}, \bibinfo {author} {\bibfnamefont {R.}~\bibnamefont {Noack}},
  \bibinfo {author} {\bibfnamefont {J.}~\bibnamefont {S\'olyom}}, \ and\
  \bibinfo {author} {\bibfnamefont {L.}~\bibnamefont {Tincani}},\ }in\
  \href@noop {} {\emph {\bibinfo {booktitle} {Computational Many-Particle
  Physics}}},\ \bibinfo {series} {Lect. Notes Phys.}, Vol.\ \bibinfo {volume}
  {739},\ \bibinfo {editor} {edited by\ \bibinfo {editor} {\bibfnamefont
  {H.}~\bibnamefont {Fehske}}, \bibinfo {editor} {\bibfnamefont
  {R.}~\bibnamefont {Schneider}}, \ and\ \bibinfo {editor} {\bibfnamefont
  {A.}~\bibnamefont {Wei\ss{}e}}}\ (\bibinfo  {publisher} {Springer},\ \bibinfo
  {address} {Berlin/Heidelberg},\ \bibinfo {year} {2008})\ pp.\ \bibinfo
  {pages} {653--664}\BibitemShut {NoStop}%
\bibitem [{\citenamefont {Marti}\ and\ \citenamefont
  {Reiher}(2010)}]{reiher10_dmrg}%
  \BibitemOpen
  \bibfield  {author} {\bibinfo {author} {\bibfnamefont {K.~H.}\ \bibnamefont
  {Marti}}\ and\ \bibinfo {author} {\bibfnamefont {M.}~\bibnamefont {Reiher}},\
  }\href@noop {} {\bibfield  {journal} {\bibinfo  {journal} {Z. Phys. Chem.}\
  }\textbf {\bibinfo {volume} {224}},\ \bibinfo {pages} {583} (\bibinfo {year}
  {2010})}\BibitemShut {NoStop}%
\bibitem [{\citenamefont {Chan}\ and\ \citenamefont
  {Sharma}(2011)}]{chanreview}%
  \BibitemOpen
  \bibfield  {author} {\bibinfo {author} {\bibfnamefont {G.~K.-L.}\
  \bibnamefont {Chan}}\ and\ \bibinfo {author} {\bibfnamefont {S.}~\bibnamefont
  {Sharma}},\ }\href@noop {} {\bibfield  {journal} {\bibinfo  {journal} {Annu.
  Rev. Phys. Chem.}\ }\textbf {\bibinfo {volume} {62}},\ \bibinfo {pages} {465}
  (\bibinfo {year} {2011})}\BibitemShut {NoStop}%
\bibitem [{\citenamefont {Tecmer}\ \emph {et~al.}(2014)\citenamefont {Tecmer},
  \citenamefont {Boguslawski}, \citenamefont {Legeza},\ and\ \citenamefont
  {Reiher}}]{C3CP53975J}%
  \BibitemOpen
  \bibfield  {author} {\bibinfo {author} {\bibfnamefont {P.}~\bibnamefont
  {Tecmer}}, \bibinfo {author} {\bibfnamefont {K.}~\bibnamefont {Boguslawski}},
  \bibinfo {author} {\bibfnamefont {{\"O}.}~\bibnamefont {Legeza}}, \ and\
  \bibinfo {author} {\bibfnamefont {M.}~\bibnamefont {Reiher}},\ }\href@noop {}
  {\bibfield  {journal} {\bibinfo  {journal} {Phys. Chem. Chem. Phys.}\ }
  (\bibinfo {year} {2014})},\ \bibinfo {note} {advance article:
  http://dx.doi.org/10.1039/C3CP53975J}\BibitemShut {NoStop}%
\bibitem [{\citenamefont {Moritz}, \citenamefont {Wolf},\ and\ \citenamefont
  {Reiher}(2005)}]{Moritz2005a}%
  \BibitemOpen
  \bibfield  {author} {\bibinfo {author} {\bibfnamefont {G.}~\bibnamefont
  {Moritz}}, \bibinfo {author} {\bibfnamefont {A.}~\bibnamefont {Wolf}}, \ and\
  \bibinfo {author} {\bibfnamefont {M.}~\bibnamefont {Reiher}},\ }\href@noop {}
  {\bibfield  {journal} {\bibinfo  {journal} {J. Chem. Phys.}\ }\textbf
  {\bibinfo {volume} {123}},\ \bibinfo {pages} {184105} (\bibinfo {year}
  {2005})}\BibitemShut {NoStop}%
\bibitem [{\citenamefont {Saue}(2011)}]{saue_primer11}%
  \BibitemOpen
  \bibfield  {author} {\bibinfo {author} {\bibfnamefont {T.}~\bibnamefont
  {Saue}},\ }\href@noop {} {\bibfield  {journal} {\bibinfo  {journal} {Chem.
  Phys. Chem.}\ }\textbf {\bibinfo {volume} {12}},\ \bibinfo {pages} {3077}
  (\bibinfo {year} {2011})}\BibitemShut {NoStop}%
\bibitem [{\citenamefont {Saue}\ and\ \citenamefont
  {Jensen}(1999)}]{saue1999_quaternion-symmetry-dirac}%
  \BibitemOpen
  \bibfield  {author} {\bibinfo {author} {\bibfnamefont {T.}~\bibnamefont
  {Saue}}\ and\ \bibinfo {author} {\bibfnamefont {H.~J.~A.}\ \bibnamefont
  {Jensen}},\ }\href@noop {} {\bibfield  {journal} {\bibinfo  {journal} {J.
  Chem. Phys.}\ }\textbf {\bibinfo {volume} {111}},\ \bibinfo {pages} {6211}
  (\bibinfo {year} {1999})}\BibitemShut {NoStop}%
\bibitem [{DIR()}]{DIRAC12}%
  \BibitemOpen
  \href@noop {} {}\bibinfo {note} {{DIRAC}, a relativistic ab initio electronic
  structure program, Release {DIRAC12} (2012), written by H.~J.~{\relax
  Aa}.~Jensen, R.~Bast, T.~Saue, and L.~Visscher, with contributions from
  V.~Bakken, K.~G.~Dyall, S.~Dubillard, U.~Ekstr{\"o}m, E.~Eliav,
  T.~Enevoldsen, T.~Fleig, O.~Fossgaard, A.~S.~P.~Gomes, T.~Helgaker,
  J.~K.~L{\ae}rdahl, Y.~S.~Lee, J.~Henriksson, M.~Ilia{\v{s}}, Ch.~R.~Jacob,
  S.~Knecht, S.~Komorovsk{\'y}, O.~Kullie, C.~V.~Larsen, H.~S.~Nataraj,
  P.~Norman, G.~Olejniczak, J.~Olsen, Y.~C.~Park, J.~K.~Pedersen,
  M.~Pernpointner, K.~Ruud, P.~Sa{\l}ek, B.~Schimmelpfennig, J.~Sikkema,
  A.~J.~Thorvaldsen, J.~Thyssen, J.~van~Stralen, S.~Villaume, O.~Visser,
  T.~Winther, and S.~Yamamoto (see http://www.diracprogram.org)}\BibitemShut
  {NoStop}%
\bibitem [{\citenamefont {Thyssen}(2001)}]{thyssendiss}%
  \BibitemOpen
  \bibfield  {author} {\bibinfo {author} {\bibfnamefont {J.}~\bibnamefont
  {Thyssen}},\ }\emph {\bibinfo {title} {Development and {A}pplications of
  {M}ethods for {C}orrelated {R}elativistic {C}alculations of {M}olecular
  {P}roperties}},\ \href@noop {} {\bibinfo {type} {Dissertation}},\ \bibinfo
  {school} {Department of Chemistry, University of Southern Denmark} (\bibinfo
  {year} {2001})\BibitemShut {NoStop}%
\bibitem [{\citenamefont {Legeza}\ \emph {et~al.}(2013)\citenamefont {Legeza},
  \citenamefont {Rohwedder}, \citenamefont {Schneider},\ and\ \citenamefont
  {Szalay}}]{new_review}%
  \BibitemOpen
  \bibfield  {author} {\bibinfo {author} {\bibfnamefont {{\"O}.}~\bibnamefont
  {Legeza}}, \bibinfo {author} {\bibfnamefont {T.}~\bibnamefont {Rohwedder}},
  \bibinfo {author} {\bibfnamefont {R.}~\bibnamefont {Schneider}}, \ and\
  \bibinfo {author} {\bibfnamefont {S.}~\bibnamefont {Szalay}},\ }\href@noop {}
  {\bibfield  {journal} {\bibinfo  {journal} {arxiv:1310.2736}\ } (\bibinfo
  {year} {2013})}\BibitemShut {NoStop}%
\bibitem [{\citenamefont {Legeza}, \citenamefont {R\"oder},\ and\ \citenamefont
  {Hess}(2003)}]{legeza_dbss2}%
  \BibitemOpen
  \bibfield  {author} {\bibinfo {author} {\bibfnamefont {{\"O}.}~\bibnamefont
  {Legeza}}, \bibinfo {author} {\bibfnamefont {J.}~\bibnamefont {R\"oder}}, \
  and\ \bibinfo {author} {\bibfnamefont {B.~A.}\ \bibnamefont {Hess}},\
  }\href@noop {} {\bibfield  {journal} {\bibinfo  {journal} {Phys. Rev. B}\
  }\textbf {\bibinfo {volume} {67}},\ \bibinfo {pages} {125114} (\bibinfo
  {year} {2003})}\BibitemShut {NoStop}%
\bibitem [{\citenamefont {Schollw\"{o}ck}(2011)}]{Schollwock2010}%
  \BibitemOpen
  \bibfield  {author} {\bibinfo {author} {\bibfnamefont {U.}~\bibnamefont
  {Schollw\"{o}ck}},\ }\href@noop {} {\bibfield  {journal} {\bibinfo  {journal}
  {Ann. Phys.}\ }\textbf {\bibinfo {volume} {326}},\ \bibinfo {pages} {96}
  (\bibinfo {year} {2011})}\BibitemShut {NoStop}%
\bibitem [{\citenamefont {Legeza}\ and\ \citenamefont
  {S\'olyom}(2004)}]{legeza_dbss3}%
  \BibitemOpen
  \bibfield  {author} {\bibinfo {author} {\bibfnamefont {{\"O}.}~\bibnamefont
  {Legeza}}\ and\ \bibinfo {author} {\bibfnamefont {J.}~\bibnamefont
  {S\'olyom}},\ }\href@noop {} {\bibfield  {journal} {\bibinfo  {journal}
  {Phys. Rev. B}\ }\textbf {\bibinfo {volume} {70}},\ \bibinfo {pages} {205118}
  (\bibinfo {year} {2004})}\BibitemShut {NoStop}%
\bibitem [{\citenamefont {Schollw\"ock}(2005)}]{schollwoeck}%
  \BibitemOpen
  \bibfield  {author} {\bibinfo {author} {\bibfnamefont {U.}~\bibnamefont
  {Schollw\"ock}},\ }\href@noop {} {\bibfield  {journal} {\bibinfo  {journal}
  {Rev.\ Mod.\ Phys.}\ }\textbf {\bibinfo {volume} {77}},\ \bibinfo {pages}
  {259} (\bibinfo {year} {2005})}\BibitemShut {NoStop}%
\bibitem [{\citenamefont {Moritz}\ and\ \citenamefont
  {Reiher}(2006)}]{Moritz2006}%
  \BibitemOpen
  \bibfield  {author} {\bibinfo {author} {\bibfnamefont {G.}~\bibnamefont
  {Moritz}}\ and\ \bibinfo {author} {\bibfnamefont {M.}~\bibnamefont
  {Reiher}},\ }\href@noop {} {\bibfield  {journal} {\bibinfo  {journal} {J.
  Chem. Phys.}\ }\textbf {\bibinfo {volume} {124}},\ \bibinfo {pages} {034103}
  (\bibinfo {year} {2006})}\BibitemShut {NoStop}%
\bibitem [{\citenamefont {Legeza}\ and\ \citenamefont
  {F\'ath}(1996)}]{legeza1996}%
  \BibitemOpen
  \bibfield  {author} {\bibinfo {author} {\bibfnamefont {{\"O}.}~\bibnamefont
  {Legeza}}\ and\ \bibinfo {author} {\bibfnamefont {G.}~\bibnamefont
  {F\'ath}},\ }\href@noop {} {\bibfield  {journal} {\bibinfo  {journal} {Phys.
  Rev. B}\ }\textbf {\bibinfo {volume} {53}},\ \bibinfo {pages} {14349}
  (\bibinfo {year} {1996})}\BibitemShut {NoStop}%
\bibitem [{\citenamefont {Legeza}\ and\ \citenamefont
  {S\'olyom}(2003)}]{legeza_dbss}%
  \BibitemOpen
  \bibfield  {author} {\bibinfo {author} {\bibfnamefont {{\"O}.}~\bibnamefont
  {Legeza}}\ and\ \bibinfo {author} {\bibfnamefont {J.}~\bibnamefont
  {S\'olyom}},\ }\href@noop {} {\bibfield  {journal} {\bibinfo  {journal}
  {Phys. Rev. B}\ }\textbf {\bibinfo {volume} {68}},\ \bibinfo {pages} {195116}
  (\bibinfo {year} {2003})}\BibitemShut {NoStop}%
\bibitem [{\citenamefont {Legeza}\ and\ \citenamefont
  {S\'olyom}()}]{legeza2004-leiden}%
  \BibitemOpen
  \bibfield  {author} {\bibinfo {author} {\bibfnamefont {{\"O}.}~\bibnamefont
  {Legeza}}\ and\ \bibinfo {author} {\bibfnamefont {J.}~\bibnamefont
  {S\'olyom}},\ }\href@noop {} {\bibinfo  {journal} {International Workshop on
  ``Recent Progress and Prospects in Density-Matrix Renormalization``, Lorentz
  Center, Leiden University, The Netherlands,
  http://www.itp.uni-hannover.de/$\sim$jeckelm/dmrg/workshop/proceedings/}\
  }\BibitemShut {NoStop}%
\bibitem [{\citenamefont {Barcza}\ \emph {et~al.}(2011)\citenamefont {Barcza},
  \citenamefont {Legeza}, \citenamefont {Marti},\ and\ \citenamefont
  {Reiher}}]{dmrg-quantum-information-analysis-2011}%
  \BibitemOpen
\bibfield  {journal} {  }\bibfield  {author} {\bibinfo {author} {\bibfnamefont
  {G.}~\bibnamefont {Barcza}}, \bibinfo {author} {\bibfnamefont
  {{\"O}.}~\bibnamefont {Legeza}}, \bibinfo {author} {\bibfnamefont {K.~H.}\
  \bibnamefont {Marti}}, \ and\ \bibinfo {author} {\bibfnamefont
  {M.}~\bibnamefont {Reiher}},\ }\href@noop {} {\bibfield  {journal} {\bibinfo
  {journal} {Phys. Rev. A}\ }\textbf {\bibinfo {volume} {83}},\ \bibinfo
  {pages} {012508} (\bibinfo {year} {2011})}\BibitemShut {NoStop}%
\bibitem [{\citenamefont {Moritz}, \citenamefont {Hess},\ and\ \citenamefont
  {Reiher}(2005)}]{dmrggerrit3}%
  \BibitemOpen
  \bibfield  {author} {\bibinfo {author} {\bibfnamefont {G.}~\bibnamefont
  {Moritz}}, \bibinfo {author} {\bibfnamefont {B.}~\bibnamefont {Hess}}, \ and\
  \bibinfo {author} {\bibfnamefont {M.}~\bibnamefont {Reiher}},\ }\href@noop {}
  {\bibfield  {journal} {\bibinfo  {journal} {J. Chem. Phys.}\ }\textbf
  {\bibinfo {volume} {122}},\ \bibinfo {pages} {024107} (\bibinfo {year}
  {2005})}\BibitemShut {NoStop}%
\bibitem [{\citenamefont {Barcza}\ \emph {et~al.}(2013)\citenamefont {Barcza},
  \citenamefont {Legeza}, \citenamefont {Noack},\ and\ \citenamefont
  {S\'olyom}}]{barcza2012}%
  \BibitemOpen
  \bibfield  {author} {\bibinfo {author} {\bibfnamefont {G.}~\bibnamefont
  {Barcza}}, \bibinfo {author} {\bibfnamefont {{\"O}.}~\bibnamefont {Legeza}},
  \bibinfo {author} {\bibfnamefont {M.~R.}\ \bibnamefont {Noack}}, \ and\
  \bibinfo {author} {\bibfnamefont {J.}~\bibnamefont {S\'olyom}},\ }\href@noop
  {} {\bibfield  {journal} {\bibinfo  {journal} {Phys. Rev. B}\ }\textbf
  {\bibinfo {volume} {86}},\ \bibinfo {pages} {075133} (\bibinfo {year}
  {2013})}\BibitemShut {NoStop}%
\bibitem [{\citenamefont {Seijo}(1995)}]{seijo1995_tlh}%
  \BibitemOpen
  \bibfield  {author} {\bibinfo {author} {\bibfnamefont {L.}~\bibnamefont
  {Seijo}},\ }\href@noop {} {\bibfield  {journal} {\bibinfo  {journal} {J.
  Chem. Phys.}\ }\textbf {\bibinfo {volume} {102}},\ \bibinfo {pages} {8078}
  (\bibinfo {year} {1995})}\BibitemShut {NoStop}%
\bibitem [{\citenamefont {van Lenthe}, \citenamefont {Snijders},\ and\
  \citenamefont {Baerends}(1996)}]{lenthe1996_tlh}%
  \BibitemOpen
  \bibfield  {author} {\bibinfo {author} {\bibfnamefont {E.}~\bibnamefont {van
  Lenthe}}, \bibinfo {author} {\bibfnamefont {J.~G.}\ \bibnamefont {Snijders}},
  \ and\ \bibinfo {author} {\bibfnamefont {E.~J.}\ \bibnamefont {Baerends}},\
  }\href@noop {} {\bibfield  {journal} {\bibinfo  {journal} {J. Chem. Phys.}\
  }\textbf {\bibinfo {volume} {105}},\ \bibinfo {pages} {6505} (\bibinfo {year}
  {1996})}\BibitemShut {NoStop}%
\bibitem [{\citenamefont {Han}, \citenamefont {Bae},\ and\ \citenamefont
  {Lee}(1999)}]{han1999_tlh}%
  \BibitemOpen
  \bibfield  {author} {\bibinfo {author} {\bibfnamefont {Y.-K.}\ \bibnamefont
  {Han}}, \bibinfo {author} {\bibfnamefont {C.}~\bibnamefont {Bae}}, \ and\
  \bibinfo {author} {\bibfnamefont {Y.~S.}\ \bibnamefont {Lee}},\ }\href
  {\doibase 10.1063/1.478901} {\bibfield  {journal} {\bibinfo  {journal} {J.
  Chem. Phys.}\ }\textbf {\bibinfo {volume} {110}},\ \bibinfo {pages} {9353}
  (\bibinfo {year} {1999})}\BibitemShut {NoStop}%
\bibitem [{\citenamefont {Titov}\ \emph {et~al.}(2000)\citenamefont {Titov},
  \citenamefont {Mosyagin}, \citenamefont {Alekseyev},\ and\ \citenamefont
  {Buenker}}]{titov2000_tlh}%
  \BibitemOpen
  \bibfield  {author} {\bibinfo {author} {\bibfnamefont {A.~V.}\ \bibnamefont
  {Titov}}, \bibinfo {author} {\bibfnamefont {N.~S.}\ \bibnamefont {Mosyagin}},
  \bibinfo {author} {\bibfnamefont {A.~B.}\ \bibnamefont {Alekseyev}}, \ and\
  \bibinfo {author} {\bibfnamefont {R.~J.}\ \bibnamefont {Buenker}},\
  }\href@noop {} {\bibfield  {journal} {\bibinfo  {journal} {Int. J. Quant.
  Chem.}\ }\textbf {\bibinfo {volume} {81}},\ \bibinfo {pages} {409} (\bibinfo
  {year} {2000})}\BibitemShut {NoStop}%
\bibitem [{\citenamefont {F{\ae}gri}\ and\ \citenamefont
  {Visscher}(2001)}]{visscher2001_tlh_ccsdt}%
  \BibitemOpen
  \bibfield  {author} {\bibinfo {author} {\bibfnamefont {K.}~\bibnamefont
  {F{\ae}gri}}\ and\ \bibinfo {author} {\bibfnamefont {L.}~\bibnamefont
  {Visscher}},\ }\href@noop {} {\bibfield  {journal} {\bibinfo  {journal}
  {Theoret. Chem. Acc.}\ }\textbf {\bibinfo {volume} {105}},\ \bibinfo {pages}
  {265} (\bibinfo {year} {2001})}\BibitemShut {NoStop}%
\bibitem [{\citenamefont {Mayer}, \citenamefont {Kr\"uger},\ and\ \citenamefont
  {R\"osch}(2001)}]{mayer2001_tlh}%
  \BibitemOpen
  \bibfield  {author} {\bibinfo {author} {\bibfnamefont {M.}~\bibnamefont
  {Mayer}}, \bibinfo {author} {\bibfnamefont {S.}~\bibnamefont {Kr\"uger}}, \
  and\ \bibinfo {author} {\bibfnamefont {N.}~\bibnamefont {R\"osch}},\
  }\href@noop {} {\bibfield  {journal} {\bibinfo  {journal} {J. Chem. Phys.}\
  }\textbf {\bibinfo {volume} {115}},\ \bibinfo {pages} {4411} (\bibinfo {year}
  {2001})}\BibitemShut {NoStop}%
\bibitem [{\citenamefont {Choi}, \citenamefont {Han},\ and\ \citenamefont
  {Lee}(2001)}]{choi2001_tlh}%
  \BibitemOpen
  \bibfield  {author} {\bibinfo {author} {\bibfnamefont {Y.~J.}\ \bibnamefont
  {Choi}}, \bibinfo {author} {\bibfnamefont {Y.-K.}\ \bibnamefont {Han}}, \
  and\ \bibinfo {author} {\bibfnamefont {Y.~S.}\ \bibnamefont {Lee}},\ }\href
  {\doibase 10.1063/1.1389289} {\bibfield  {journal} {\bibinfo  {journal} {J.
  Chem. Phys.}\ }\textbf {\bibinfo {volume} {115}},\ \bibinfo {pages} {3448}
  (\bibinfo {year} {2001})}\BibitemShut {NoStop}%
\bibitem [{\citenamefont {Ilias}\ \emph {et~al.}(2001)\citenamefont {Ilias},
  \citenamefont {Kell\"o}, \citenamefont {Visscher},\ and\ \citenamefont
  {Schimmelpfennig}}]{ilias2001_tlh}%
  \BibitemOpen
  \bibfield  {author} {\bibinfo {author} {\bibfnamefont {M.}~\bibnamefont
  {Ilias}}, \bibinfo {author} {\bibfnamefont {V.}~\bibnamefont {Kell\"o}},
  \bibinfo {author} {\bibfnamefont {L.}~\bibnamefont {Visscher}}, \ and\
  \bibinfo {author} {\bibfnamefont {B.}~\bibnamefont {Schimmelpfennig}},\
  }\href {\doibase 10.1063/1.1413510} {\bibfield  {journal} {\bibinfo
  {journal} {J. Chem. Phys.}\ }\textbf {\bibinfo {volume} {115}},\ \bibinfo
  {pages} {9667} (\bibinfo {year} {2001})}\BibitemShut {NoStop}%
\bibitem [{\citenamefont {Choi}\ and\ \citenamefont
  {Lee}(2003)}]{choi2003_tlh}%
  \BibitemOpen
  \bibfield  {author} {\bibinfo {author} {\bibfnamefont {Y.~J.}\ \bibnamefont
  {Choi}}\ and\ \bibinfo {author} {\bibfnamefont {Y.~S.}\ \bibnamefont {Lee}},\
  }\href@noop {} {\bibfield  {journal} {\bibinfo  {journal} {J. Chem. Phys.}\
  }\textbf {\bibinfo {volume} {119}},\ \bibinfo {pages} {2014} (\bibinfo {year}
  {2003})}\BibitemShut {NoStop}%
\bibitem [{\citenamefont {Zeng}, \citenamefont {Fedorov},\ and\ \citenamefont
  {Klobukowski}(2010)}]{zeng2010-tlh-so}%
  \BibitemOpen
  \bibfield  {author} {\bibinfo {author} {\bibfnamefont {T.}~\bibnamefont
  {Zeng}}, \bibinfo {author} {\bibfnamefont {D.~G.}\ \bibnamefont {Fedorov}}, \
  and\ \bibinfo {author} {\bibfnamefont {M.}~\bibnamefont {Klobukowski}},\
  }\href@noop {} {\bibfield  {journal} {\bibinfo  {journal} {J. Chem. Phys.}\
  }\textbf {\bibinfo {volume} {132}},\ \bibinfo {pages} {074102} (\bibinfo
  {year} {2010})}\BibitemShut {NoStop}%
\bibitem [{\citenamefont {Dyall}(2002)}]{dyall_6p_all}%
  \BibitemOpen
  \bibfield  {author} {\bibinfo {author} {\bibfnamefont {K.~G.}\ \bibnamefont
  {Dyall}},\ }\href@noop {} {\bibfield  {journal} {\bibinfo  {journal} {Theor.
  Chem. Acc.}\ }\textbf {\bibinfo {volume} {108}},\ \bibinfo {pages} {335}
  (\bibinfo {year} {2002})},\ \bibinfo {note} {{E}rratum: Theor. Chem. Acc.
  (2003) 109:284; Revision: Theor. Chem. Acc. (2006) 115:441.}\BibitemShut
  {Stop}%
\bibitem [{\citenamefont {Dyall}(2012)}]{dyall_6p_core_corr2012}%
  \BibitemOpen
  \bibfield  {author} {\bibinfo {author} {\bibfnamefont {K.~G.}\ \bibnamefont
  {Dyall}},\ }\href@noop {} {\bibfield  {journal} {\bibinfo  {journal} {Theor.
  Chem. Acc.}\ }\textbf {\bibinfo {volume} {131}},\ \bibinfo {pages} {1217}
  (\bibinfo {year} {2012})}\BibitemShut {NoStop}%
\bibitem [{\citenamefont {{Dunning Jr.}}(1989)}]{dunning89}%
  \BibitemOpen
  \bibfield  {author} {\bibinfo {author} {\bibfnamefont {T.~H.}\ \bibnamefont
  {{Dunning Jr.}}},\ }\href@noop {} {\bibfield  {journal} {\bibinfo  {journal}
  {J.\,Chem.\,Phys.}\ }\textbf {\bibinfo {volume} {90}},\ \bibinfo {pages}
  {1007} (\bibinfo {year} {1989})}\BibitemShut {NoStop}%
\bibitem [{\citenamefont {Legeza}()}]{Legeza}%
  \BibitemOpen
  \bibfield  {author} {\bibinfo {author} {\bibfnamefont {{\"O}.}~\bibnamefont
  {Legeza}},\ }\href@noop {} {\emph {\bibinfo {title}
  {\textsc{QC-DMRG-Budapest}, A Program for Quantum Chemical {DMRG}
  Calculations. { \rm Copyright 2000--2013, HAS Wigner Budapest}}}}\BibitemShut
  {NoStop}%
\bibitem [{mp2()}]{mp2-no}%
  \BibitemOpen
  \href@noop {} {}\bibinfo {note} {Option \textsc{.MP2 NO} in the wave function
  section of the program package \texttt{Dirac12}.}\BibitemShut {Stop}%
\bibitem [{\citenamefont {Fleig}, \citenamefont {Olsen},\ and\ \citenamefont
  {Visscher}(2003)}]{fleig03}%
  \BibitemOpen
  \bibfield  {author} {\bibinfo {author} {\bibfnamefont {T.}~\bibnamefont
  {Fleig}}, \bibinfo {author} {\bibfnamefont {J.}~\bibnamefont {Olsen}}, \ and\
  \bibinfo {author} {\bibfnamefont {L.}~\bibnamefont {Visscher}},\ }\href@noop
  {} {\bibfield  {journal} {\bibinfo  {journal} {J.\,Chem.\,Phys.}\ }\textbf
  {\bibinfo {volume} {119}},\ \bibinfo {pages} {2963} (\bibinfo {year}
  {2003})}\BibitemShut {NoStop}%
\bibitem [{\citenamefont {Fleig}\ \emph {et~al.}(2006)\citenamefont {Fleig},
  \citenamefont {Jensen}, \citenamefont {Olsen},\ and\ \citenamefont
  {Visscher}}]{fleig06}%
  \BibitemOpen
  \bibfield  {author} {\bibinfo {author} {\bibfnamefont {T.}~\bibnamefont
  {Fleig}}, \bibinfo {author} {\bibfnamefont {H.~J.~{\Aa}.}\ \bibnamefont
  {Jensen}}, \bibinfo {author} {\bibfnamefont {J.}~\bibnamefont {Olsen}}, \
  and\ \bibinfo {author} {\bibfnamefont {L.}~\bibnamefont {Visscher}},\
  }\href@noop {} {\bibfield  {journal} {\bibinfo  {journal} {J.\,Chem.\,Phys.}\
  }\textbf {\bibinfo {volume} {124}},\ \bibinfo {pages} {104106} (\bibinfo
  {year} {2006})}\BibitemShut {NoStop}%
\bibitem [{\citenamefont {Knecht}, \citenamefont {Jensen},\ and\ \citenamefont
  {Fleig}(2010)}]{knecht10a}%
  \BibitemOpen
  \bibfield  {author} {\bibinfo {author} {\bibfnamefont {S.}~\bibnamefont
  {Knecht}}, \bibinfo {author} {\bibfnamefont {H.~J.~{\Aa}.}\ \bibnamefont
  {Jensen}}, \ and\ \bibinfo {author} {\bibfnamefont {T.}~\bibnamefont
  {Fleig}},\ }\href@noop {} {\bibfield  {journal} {\bibinfo  {journal}
  {J.\,Chem.\,Phys.}\ }\textbf {\bibinfo {volume} {132}},\ \bibinfo {pages}
  {014108} (\bibinfo {year} {2010})}\BibitemShut {NoStop}%
\bibitem [{mrc()}]{mrcceng}%
  \BibitemOpen
  \href@noop {} {}\bibinfo {note} {{\sc Mrcc}, a string-based quantum chemical
  program suite written by M. {K\'allay}. See also Ref. \citenum{Kallay_2001}
  as well as http://www.mrcc.hu/}\BibitemShut {NoStop}%
\bibitem [{\citenamefont {K\'allay}\ and\ \citenamefont
  {Surj\'an}(2001)}]{Kallay_2001}%
  \BibitemOpen
  \bibfield  {author} {\bibinfo {author} {\bibfnamefont {M.}~\bibnamefont
  {K\'allay}}\ and\ \bibinfo {author} {\bibfnamefont {P.~R.}\ \bibnamefont
  {Surj\'an}},\ }\href@noop {} {\bibfield  {journal} {\bibinfo  {journal} {J.
  Chem. Phys.}\ }\textbf {\bibinfo {volume} {115}},\ \bibinfo {pages} {2945}
  (\bibinfo {year} {2001})}\BibitemShut {NoStop}%
\bibitem [{\citenamefont {Boguslawski}\ \emph {et~al.}(2012)\citenamefont
  {Boguslawski}, \citenamefont {Tecmer}, \citenamefont {Legeza},\ and\
  \citenamefont {Reiher}}]{dmrg-entanglement12}%
  \BibitemOpen
  \bibfield  {author} {\bibinfo {author} {\bibfnamefont {K.}~\bibnamefont
  {Boguslawski}}, \bibinfo {author} {\bibfnamefont {P.}~\bibnamefont {Tecmer}},
  \bibinfo {author} {\bibfnamefont {{\"O}.}~\bibnamefont {Legeza}}, \ and\
  \bibinfo {author} {\bibfnamefont {M.}~\bibnamefont {Reiher}},\ }\href@noop {}
  {\bibfield  {journal} {\bibinfo  {journal} {J. Phys. Chem. Lett.}\ }\textbf
  {\bibinfo {volume} {3}},\ \bibinfo {pages} {3129} (\bibinfo {year}
  {2012})}\BibitemShut {NoStop}%
\bibitem [{\citenamefont {Boguslawski}\ \emph {et~al.}(2013)\citenamefont
  {Boguslawski}, \citenamefont {Tecmer}, \citenamefont {Barcza}, \citenamefont
  {Legeza},\ and\ \citenamefont {Reiher}}]{boguslawski2013a}%
  \BibitemOpen
  \bibfield  {author} {\bibinfo {author} {\bibfnamefont {K.}~\bibnamefont
  {Boguslawski}}, \bibinfo {author} {\bibfnamefont {P.}~\bibnamefont {Tecmer}},
  \bibinfo {author} {\bibfnamefont {G.}~\bibnamefont {Barcza}}, \bibinfo
  {author} {\bibfnamefont {{\"O}.}~\bibnamefont {Legeza}}, \ and\ \bibinfo
  {author} {\bibfnamefont {M.}~\bibnamefont {Reiher}},\ }\href {\doibase
  10.1021/ct400247p} {\bibfield  {journal} {\bibinfo  {journal} {J. Chem.
  Theory Comput.}\ }\textbf {\bibinfo {volume} {9}},\ \bibinfo {pages} {2959}
  (\bibinfo {year} {2013})}\BibitemShut {NoStop}%
\bibitem [{\citenamefont {Kallay}\ \emph {et~al.}(2011)\citenamefont {Kallay},
  \citenamefont {Nataraj}, \citenamefont {Sahoo}, \citenamefont {Das},\ and\
  \citenamefont {Visscher}}]{kallay2011_4cCC}%
  \BibitemOpen
  \bibfield  {author} {\bibinfo {author} {\bibfnamefont {M.}~\bibnamefont
  {Kallay}}, \bibinfo {author} {\bibfnamefont {H.}~\bibnamefont {Nataraj}},
  \bibinfo {author} {\bibfnamefont {B.}~\bibnamefont {Sahoo}}, \bibinfo
  {author} {\bibfnamefont {B.}~\bibnamefont {Das}}, \ and\ \bibinfo {author}
  {\bibfnamefont {L.}~\bibnamefont {Visscher}},\ }\href@noop {} {\bibfield
  {journal} {\bibinfo  {journal} {Phys. Rev. A}\ }\textbf {\bibinfo {volume}
  {83}},\ \bibinfo {pages} {1} (\bibinfo {year} {2011})}\BibitemShut {NoStop}%
\bibitem [{\citenamefont {Grundstr\"om}\ and\ \citenamefont
  {Valberg}(1938)}]{tlh_experiment_I_1938}%
  \BibitemOpen
  \bibfield  {author} {\bibinfo {author} {\bibfnamefont {B.}~\bibnamefont
  {Grundstr\"om}}\ and\ \bibinfo {author} {\bibfnamefont {P.}~\bibnamefont
  {Valberg}},\ }\href {\doibase 10.1007/BF01374953} {\bibfield  {journal}
  {\bibinfo  {journal} {Z. Phys.}\ }\textbf {\bibinfo {volume} {108}},\
  \bibinfo {pages} {326} (\bibinfo {year} {1938})}\BibitemShut {NoStop}%
\bibitem [{\citenamefont {Huber}\ and\ \citenamefont
  {Herzberg}(1979)}]{huber79}%
  \BibitemOpen
  \bibinfo {editor} {\bibfnamefont {K.}~\bibnamefont {Huber}}\ and\ \bibinfo
  {editor} {\bibfnamefont {G.}~\bibnamefont {Herzberg}},\ eds.,\ \href@noop {}
  {\emph {\bibinfo {title} {Molecular Spectra and Molecular Structure Constants
  of Diatomic Molecules}}}\ (\bibinfo  {publisher} {Van Nostrand, NewYork},\
  \bibinfo {year} {1979})\BibitemShut {NoStop}%
\bibitem [{\citenamefont {Urban}\ \emph {et~al.}(1989)\citenamefont {Urban},
  \citenamefont {Bahnmaier}, \citenamefont {Magg},\ and\ \citenamefont
  {Jones}}]{Urban1989_tlh_experiment_II}%
  \BibitemOpen
  \bibfield  {author} {\bibinfo {author} {\bibfnamefont {R.-D.}\ \bibnamefont
  {Urban}}, \bibinfo {author} {\bibfnamefont {A.~H.}\ \bibnamefont
  {Bahnmaier}}, \bibinfo {author} {\bibfnamefont {U.}~\bibnamefont {Magg}}, \
  and\ \bibinfo {author} {\bibfnamefont {H.}~\bibnamefont {Jones}},\ }\href
  {\doibase 10.1016/0009-2614(89)87368-3} {\bibfield  {journal} {\bibinfo
  {journal} {Chem. Phys. Lett.}\ }\textbf {\bibinfo {volume} {158}},\ \bibinfo
  {pages} {443} (\bibinfo {year} {1989})}\BibitemShut {NoStop}%
\end{thebibliography}

\newcommand{\Aa}[0]{Aa}

\end{document}